\documentclass[journal=jpccck,manuscript=article]{achemso}
\usepackage{amsmath}
\usepackage{amssymb}
\usepackage{graphicx}
\usepackage[version=3]{mhchem}

\title{Molecular Split-Ring Resonators Based on Metal String Complexes} 
\author{Yao Shen}
\author{Hsin-Yu Ko}
\author{Qing Ai}
\author{Shie-Ming Peng}
\author{Bih-Yaw Jin }
\email{byjin@ntu.edu.tw}
\affiliation{Department of Chemistry and Center for Emerging Material
  and Advanced Devices and Center for Quantum Science and Engineering,
  National Taiwan University, Taipei 10617, Taiwan}

\def\>{\rangle}
\def\<{\langle}
\newcommand{\rttensor}[1]{\overline{\overline{#1}}}

\begin{document}
\begin{abstract}
  Metal string complexes or extended metal atom chains (EMACs) belong
  to a family of molecules that consist of a linear chain of directly
  bonded metal atoms embraced helically by four multidentate organic
  ligands. These four organic ligands are usually made up of repeating
  pyridyl units, single-nitrogen-substituted heterocyclic annulenes,
  bridged by independent amido groups. Here, in this paper, we show
  that these heterocyclic annulenes are actually nanoscale molecular
  split-ring resonators (SRRs) that can exhibit simultaneous negative
  electric permittivity and magnetic permeability in the UV-Vis
  region. Moreover, a monolayer of self-assembled EMACs is a periodic
  array of molecular SRRs which can be considered as a negative
  refractive index material. In the molecular scale, where the
  quantum-size effect is significant, we apply the tight-binding
  method to obtain the frequency-dependent permittivity and
  permeability of these molecular SRRs with their tensorial properties
  carefully considered.
\end{abstract}
\maketitle


\section{Introduction}

Negative refraction is an exotic refractive process which can be
applied on high resolution lenses to overcome optical diffraction
limits\cite{k1}, local field enhancements\cite{k2}, and
high-sensitivity detections\cite{k3,k4,k5}.  In 1968, Veselago first
investigated the electromagnetic properties of negative index material
(NIM)\cite{k6}. It was not until the late 1990s, Pendry \textit{et
  al.}\cite{k2,k7,k8} introduced the use of metamaterial which allowed
NIM to become a prosperous field of study.  Pendry \textit{et al.}
also pointed out that metamaterial can be realized by a kind of
conductive ring with a split pointing to a specific direction which is
called split-ring resonators (SRRs)\cite{k2}. After two years, it was
experimentally realized by Shelby \textit{et al.}\cite{k10}.  This
remarkable discovery makes SRR and metamaterial \cite{k9} new branch
of research, including liquid crystal magnetic control\cite{k11},
cloaking\cite{k12,k13} and toroidal dipole moments\cite{k14},
\textit{etc}.  However, metamaterials normally act on specific
wavelength ranges that depend on the sizes of building circuits. For
example, the experimental measurement of SRR is made in $5$ mm cell
dimension and operates on microwave with wavelength $3$
cm\cite{k10}. To find visible-frequency NIM, the challenge is to make
arrays of SRR in nanometer scale. During the past decade, researchers
have manufactured SRRs with sizes ranging from a few
minimeters\cite{k15} to several hundreds of
nanometers\cite{k16,k17,k18,k19,k20,k21}.  In this paper, we research
a class of one-dimensional metal string complexes, which is also
called extended metal atom chains (EMACs), for potential application
as molecular-scale SRRs (MSRRs). EMACs are linear strings of
transition metal atoms with four polypyridylamido ligands wrapping
around them. Although there exist a number of metallic nanowires
\cite{k23-0,k23-1} already, EMACs are in fact the thinnest molecular
electrical wires\cite{k23,k24,k25} that can show intriguing magnetism
and large single-molecule conductances\cite{k26,k27,k28,k29,k30}.
Since the extended ligands of EMACs are generally pyridyl chains, they
offer great configurations of well ordered arrays of conjugated rings,
which can also be modeled as Gentile oscillators of intermediate
statistics \cite{Shen13,Shen10,Shen07}. We wish to point out in this
paper that the extended ligands of EMACs are in fact a new type of
molecular SRRs. The special configuration of
polypyridylamido ligands around the central metal string in EMACs
makes the negative refraction possible. Moreover, due to the quantum
confinement of $\pi$-electrons in the pyridyl groups, EMACs as the
negative refractive index materials are promising to realize
simultaneous negativity for both electric permittivity and magnetic
permeability  in the UV-Vis region.

This paper is organized as follows: In section II, we give the theoretical
derivation of the permittivity and permeability of EMACs using a simple
tight-binding model. Three dimensional helical configuration of ligands 
are both discussed in section III. Finally, in section IV, the main results 
are concluded and future work is discussed.

\section{Theoretical Derivation}

\label{sec:threed}

\subsection{Geometry of EMACs}

\begin{figure}
\includegraphics{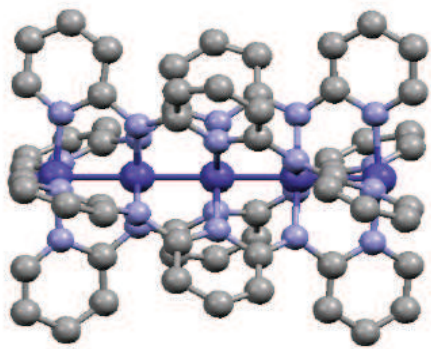}
\caption{ A linear pentanuclear chain embraced by four
  polypyridylamide ligands, $\ce{M5(tpda)_{4}}X2$
  (tpda=tripyridylamide). Axial ligands are not shown. This
  one-dimensional metal string complex consists of five transition
  metal atoms (blue) and twelve single-nitrogen-substituted
  heterocyclic annulenes as their ligands (gray). \label{fig:One} }
\end{figure}

\begin{figure}
\includegraphics[bb=0bp 0bp 300bp 160bp,clip]{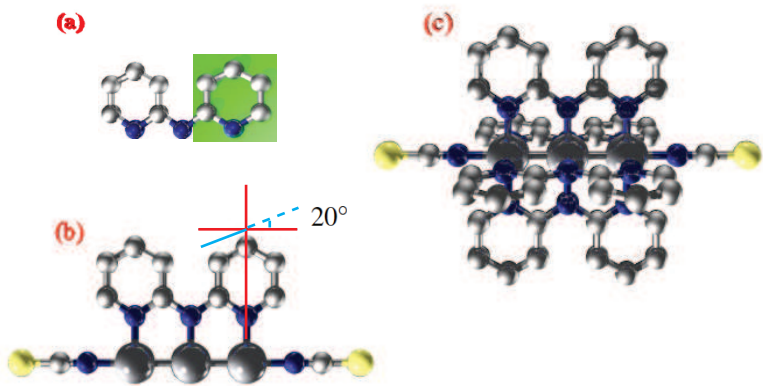}
\caption{Pyridyl units inside a trinuclear EMAC complex. The white
  balls are carbon atoms. The blue balls are nitrogen atoms. The big gray
  balls are transition metal atoms. The two yellow balls on both ends are sulphur
  atoms. (a) The green box is the pyridyl unit. (b) Two split rings
  attach to three transition metal atoms. (c) The smallest unit of
  EMACs.\label{fig:EMACS-1}}
\end{figure}

The configuration of a one-dimensional metal string complex that
consists of five transition metal atoms arranged in the linear string
with twelve pyridyl units acting as molecular SRRs helically wrapping
around them is shown in Figure~\ref{fig:One}.  In our current study,
we will treat these pyridyl units as independent quantum subsystems.
For simplicity, we focus on the shortest EMACs containing only three
transition metal atoms and eight molecular SRRs (see Figure
\ref{fig:EMACS-1}).

Figure~\ref{fig:EMACS-1}(a) shows the structure of a dipyridylamido
ligand in a trinuclear EMAC complex. The green box presents a single
heteronuclear pyridyl unit consisting of five carbon atoms (white
balls) and one nitrogen atom (blue ball). We will demonstrate later
that each pyridyl unit in an EMAC complex actually corresponds a
molecular SRR.  In Figure~\ref{fig:EMACS-1}(b), three transition metal
atoms (gray balls) serve as the backbone which holds two molecular
SRRs in a helical arrangement with a dihedral angle $\sim20^{\circ}$
between two pyridyl units that are linked by a single amido group. With
all four dipyridylamido ligands attached on the metal string with a
four-fold rotational symmetry, we get an EMAC complex
(Figure~\ref{fig:EMACS-1}(c)).

\begin{figure}
\includegraphics[bb=0bp 0bp 150bp 150bp,clip]{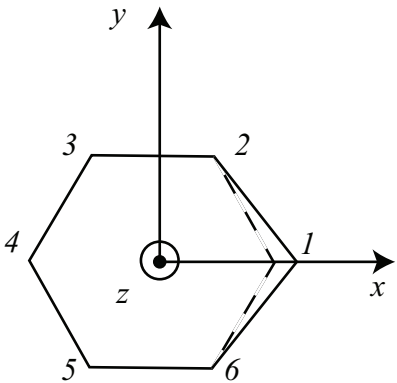} 
\caption{Schematic diagram of a pyridine molecule modeled as a
  perfect hexagon.  The origin is set at the center of the hexagon
  with the $z$ axis perpendicular to the plane. The six sites are
  labeled sequentially from $1$ to $6$, with site $1$ to be a nitrogen
  atom and others to be carbon atoms.\label{fig:two}}
\end{figure}

\subsection{H\"uckel Model}

The electromagnetic properties mainly originate from the response of
$\pi$ electrons of the ligands to the external electromagnetic
fields. We label six $p_z$ orbitals in one of the eight pyridyl groups
according to Figure~\ref{fig:two}.  Since each ligand consists of two pyridyl
units linked by an amido group, without loss of generality, we will
assume that these pyridyl units are independent and can be modeled by
the H\"uckel model as \cite{k33}
\begin{equation}
  \mathcal{H} =
  \sum_{j=1}^{6}\alpha_{j}|j\>\<j|
  +\sum_{j=1}^{6}\beta_{j,j+1}\left(|j\>\<j+1|+|j+1\>\<j|\right), 
  \label{eq:Hhu} 
\end{equation}
where $|j\>$ is the $p_{z}$-atomic orbital at site $j$ with site
energy $\alpha_{j}$, $\beta_{j,j+1}$ is the coupling strength between
$j$th and $(j+1)$th sites. Here, the cyclic condition is employed,
i.e., $|7\>=|1\>$. We assume that the only nitrogen atom is located at
site 1. Therefore, 
\begin{eqnarray}
  \alpha_{j} & = & 
  \left\{ \begin{array}{c}
      \alpha_{C}\textrm{, for }j\neq1,\\
      \alpha_{N}\textrm{, for }j=1,
    \end{array}
  \right. \\
  \beta_{j,j+1} & = & 
  \left\{ \begin{array}{c}
      \beta_{CC}\textrm{, for }j\neq1,6,\\
      \beta_{CN}\textrm{, for }j=1,6.
    \end{array}
  \right.
\end{eqnarray}
The H\"uckel Hamiltonian (\ref{eq:Hhu}) can be diagonalized numerically
as 
\begin{eqnarray}
\mathcal{H} & = & \sum_{k=1}^{6}\varepsilon_{k}|\psi_{k}\>\<\psi_{k}|,\\
|\psi_{k}\> & = & \sum_{j=1}^{6}c_{kj}|j\>.
\end{eqnarray}

\begin{figure}
\includegraphics[bb=0bp 0bp 362bp 186bp,clip]{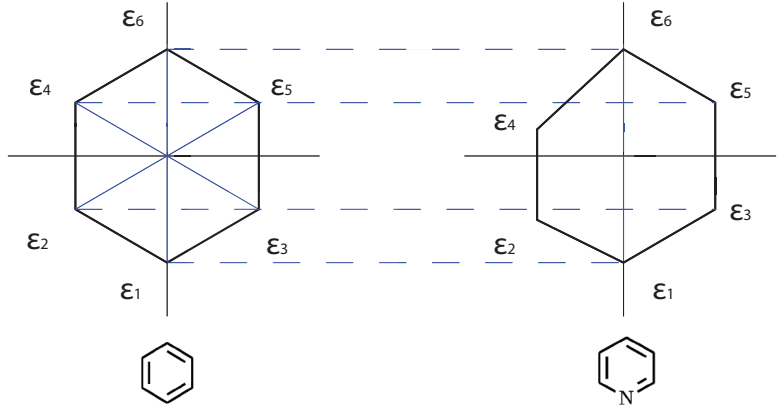}
\caption{The spectra of (left) benzene and (right) pyridine
  molecules. Due to the symmetry, there are two sets of two-fold
  degenerate states for a benzene molecule, i.e., $|\psi_{2}\>$ and
  $|\psi_{3}\>$, and $|\psi_{4}\>$ and $|\psi_{5}\>$. Because a carbon
  atom is substituted by a nitrogen atom, the symmetry is broken and
  thus there are no degenerate states for a pyridine
  molecule.\label{fig:three}}
\end{figure}

Each pyridyl unit contains six non-interacting $\pi$ electrons. On
account of the spin degrees of freedom, the ground state is the state
with all six electrons filling the three lowest molecular orbitals
(see Figure \ref{fig:three}), i.e., $|\Psi_{0}\> =
a_{1\uparrow}^{\dagger}a_{1\downarrow}^{\dagger}
a_{2\uparrow}^{\dagger}a_{2\downarrow}^{\dagger}
a_{3\uparrow}^{\dagger}a_{3\downarrow}^{\dagger}|\text{vac}\>$, 
with the total $\pi$ electron energy $E_{0} =
2(\varepsilon_{1}+\varepsilon_{2}+\varepsilon_{3})$.  Here
$a_{k\uparrow(\downarrow)}^{\dagger}$ creates an electron in the molecular orbital
$|\psi_{k}\>$ with spin up(down) and $|\text{vac}\>$ corresponds to the
vacuum state.

Within the independent electron approximation, the ground state is
coupled, through the external electromagnetic field, to single excitation
states by promoting one electron from occupied to unoccupied molecular
orbitals. There are eighteen single excitation states $|\Psi_{I}\> =
|\Psi_{a\sigma}^{r\sigma}\> =
a_{r\sigma}^{\dagger}a_{a\sigma}|\Psi_{0}\>$, $a=1,2,3$, $r=4,5,6$ and
$\sigma = \uparrow,\downarrow$, with eigen-energies
$E_{I}=E(\Psi_{a\sigma}^{r\sigma})=E_{0}+\varepsilon_{r}-\varepsilon_{a}$.
Therefore, for six non-interacting $\pi$ electrons in the absence of
applied electromagnetic field, the Hamiltonian reads
\begin{equation}
  H_{0}=\sum_{I=0}^{18}E_{I}|\Psi_{I}\>\<\Psi_{I}|.
\end{equation}

\subsection{Perturbation Theory in Rotating Frame}

In the presence of a time-dependent electromagnetic field, the total
Hamiltonian including the interaction between the electrons and the
field is 
\begin{eqnarray}
  H & = & H_{0}-\vec{\mu}\cdot\vec{E}(\vec{r},t)-\vec{m}\cdot\vec{B}(\vec{r},t)\nonumber \\
  & = & H_{0}-\vec{\mu}\cdot\vec{E}_{0}\cos(\vec{k}\cdot\vec{r}-\omega t)-\vec{m}\cdot\vec{B}_{0}\cos(\vec{k}\cdot\vec{r}-\omega t)\nonumber \\
  & \simeq & H_{0}-\vec{\mu}\cdot\vec{E}_{0}\cos(\omega t)-\vec{m}\cdot\vec{B}_{0}\cos(\omega t),
\end{eqnarray}
where $\vec{\mu}$ and $\vec{m}$ are the electric and magnetic dipole
moments, respectively. In the last line we have assumed the wavelength
of the field is much larger than the size of the molecule, i.e., $\vec{k}\cdot\vec{r}\simeq0$
if the origin of the coordinate is chosen at the center of the molecule.

Then, after transforming to the rotating frame defined by a unitary
transformation, $U^{\dagger}=\exp(-i\omega t|\Psi_{0}\>\<\Psi_{0}|\,
)$, and we have used the rotating wave approximation, the effective Hamiltonian becomes 
\begin{eqnarray}
  H^{'} & = & U^{\dagger}HU+i\dot{U}^{\dagger}U\nonumber \\
  & \simeq & \sum_{I=1}^{18}E_{k}|\Psi_{I}\>\<\Psi_{I}|+(E_{0}+\omega)|\Psi_{0}\>\<\Psi_{0}|+H_{1}^{\prime},
\end{eqnarray}
where 
\begin{eqnarray}
  H_{1}^{\prime} & = & -\frac{1}{2}\sum_{I=1}^{18}\left(\vec{\mu}_{I0}\cdot\vec{E}_{0}|\Psi_{I}\>\<\Psi_{0}|+\vec{\mu}_{0I}\cdot\vec{E}_{0}|\Psi_{0}\>\<\Psi_{I}|\right)\nonumber \\
  &  & -\frac{1}{2}\sum_{I=1}^{18}\left(\vec{m}_{I0}\cdot\vec{B}_{0}|\Psi_{I}\>\<\Psi_{0}|+\vec{m}_{0I}\cdot\vec{B}_{0}|\Psi_{0}\>\<\Psi_{I}|\right),
\end{eqnarray}
where $  \vec{\mu}_{II^{\prime}}  =
\<\Psi_{I}|\vec{\mu}|\Psi_{I^{\prime}}\>$ and 
$ \vec{m}_{II^{\prime}}  = \<\Psi_{I}|\vec{m}|\Psi_{I^{\prime}}\>$.
Notice that due to the transformation, the wave function $|\Psi\>$
and operator $A$ in the Schr\"odinger picture are transformed
according to  $|\Psi^{\prime}\> =  U^{\dagger}|\Psi\>$
and $A^{\prime}  = U^{\dagger}AU$.
Thus, the dipole operator in the rotating frame reads 
\begin{eqnarray*}
\vec{\mu^{\prime}} & = & \sum_{I=1}^{18}\left(\vec{\mu}_{I0}e^{i\omega t}|\Psi_{I}\>\<\Psi_{0}|+\vec{\mu}_{0I}e^{-i\omega t}|\Psi_{0}\>\<\Psi_{I}|\right).
\end{eqnarray*}
The ground state of the effective Hamiltonian, $H'$, is   
\begin{equation}
|\Psi_{0}^{\prime}\>=|\Psi_{0}\>+\sum_{I=1}^{18}\frac{\<\Psi_{I}|H_{1}^{\prime}|\Psi_{0}\>}{E_{0}+\omega-E_{I}}|\Psi_{I}\>.
\end{equation}
The expectation value of the dipole operator  for the ground state is 
\begin{equation}
\<\Psi_{0}^{\prime}|\vec{\mu}^{\prime}|\Psi_{0}^{\prime}\>=-\textrm{Re}\sum_{I=1}^{18}\frac{\vec{\mu}_{I0}\cdot\vec{E}_{0}}{E_{0}+\omega-E_{I}}\vec{\mu}_{0I}e^{-i\omega t}.
\end{equation}

If there are  $N$ identical molecules but each one with a different
orientation in the total volume $V$, the total electric displacement
field is 
\begin{eqnarray}
\vec{D} & = & \rttensor{\epsilon}_{0}\vec{E}_{0}+\frac{\vec{P}}{V}\nonumber \\
 & = & \rttensor{\epsilon}_{0}\vec{E}_{0}-\frac{1}{V}\sum_{n=1}^{N}\sum_{I=1}^{18}\frac{\left[\vec{\mu}_{I0}(n)\cdot\vec{E}_{0}\right]\vec{\mu}_{0I}(n)}{\left(E_{0}+\omega-E_{I}\right)}\nonumber \\
 & \equiv & \rttensor{\epsilon}\vec{E}_{0},
\end{eqnarray}
where 
\begin{eqnarray}
\epsilon_{ij} & \equiv & \epsilon_{0}\delta_{ij}-\frac{1}{V}\sum_{n=1}^{N}\sum_{I=1}^{18}\frac{\mu_{0I}^{(i)}(n)\mu_{I0}^{(j)}(n)}{\left(E_{0}+\omega-E_{I}\right)},\quad\textrm{ for }\quad i,j=x,y,z\\
\vec{\mu}_{II^{\prime}}(n) & = & \mu_{II^{\prime}}^{(x)}(n)\hat{e}_{x}+\mu_{II^{\prime}}^{(y)}(n)\hat{e}_{y}+\mu_{II^{\prime}}^{(z)}(n)\hat{e}_{z},
\end{eqnarray}
and $\vec{\mu}(n)$ is the $n$th molecular electric dipole operator,
$\hat{e}_{i}$ is the unit vector of the lab coordinate system.

\begin{figure}
\includegraphics[bb=0bp 0bp 244bp 150bp,clip,scale=0.7]{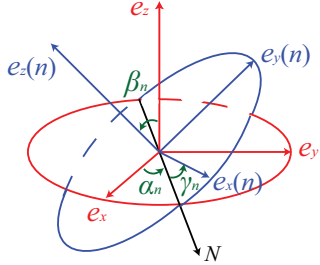} 
\caption{The matrix $R(\alpha_{n},\beta_{n},\gamma_{n})$ corresponds to the
rotation from the lab coordinate system $(\hat{e}_{x},\hat{e}_{y},\hat{e}_{z})$
to the $n$th molecule coordinate system $(\hat{e}_{x}(n),\hat{e}_{y}(n),\hat{e}_{z}(n))$.
$\hat{N}$ is the line of nodes. $\alpha_{n}$, $\beta_{n}$, $\gamma_{n}$
are respectively the angle between $\hat{e}_{x}$ and $\hat{N}$,
the angle between $\hat{e}_{z}$ and $\hat{e}_{z}(n)$, and the angle
between $\hat{N}$ and $\hat{e}_{x}(n)$.\label{fig:ro}}
\end{figure}

In order to obtain the relative dielectric constant of the material,
we have to transform back to the lab coordinate system. The relation
between the electric field in the molecular coordinate system
$\vec{E}^{M}$ and the lab coordinate system $\vec{E}^{L}$ is (see
Figure \ref{fig:ro}) 
\begin{equation}
\vec{E}^{L}=R^{-1}(\alpha_{n},\beta_{n},\gamma_{n})\vec{E}^{M},
\end{equation}
where 
\begin{eqnarray}
R(\alpha_{n},\beta_{n},\gamma_{n}) & = & R_{z}(\gamma_{n})R_{x}(\beta_{n})R_{z}(\alpha_{n}),\\
R_{x}(\beta_{n}) & = & \begin{pmatrix}1 & 0 & 0\\
0 & \cos\beta_{n} & -\sin\beta_{n}\\
0 & \sin\beta_{n} & \cos\beta_{n}
\end{pmatrix},\\
R_{z}(\alpha_{n}) & = & \begin{pmatrix}\cos\alpha_{n} & -\sin\alpha_{n} & 0\\
\sin\alpha_{n} & \cos\alpha_{n} & 0\\
0 & 0 & 1
\end{pmatrix},
\end{eqnarray}
are the rotations around $x$($z$) axis with an angle $\beta_{n}$($\alpha_{n}$),
$\alpha_{n}$, $\beta_{n}$, $\gamma_{n}$ are respectively the angle
between $\hat{e}_{x}$ and $\hat{N}$, the angle between $\hat{e}_{z}$
and $\hat{e}_{z}(n)$, and the angle between $\hat{N}$ and $\hat{e}_{x}(n)$,
$\hat{N}$ is the line of nodes. Finally, the relative dielectric constant
of the material in the lab coordinate system is 
\begin{equation}
\epsilon_{ij}^{r}\equiv\delta_{ij}-\sum_{n=1}^{N}\sum_{k=1}^{18}\frac{\left[R^{-1}(\alpha_{n},\beta_{n},\gamma_{n})\vec{\mu}_{0I}(n)\right]\cdot\hat{e}_{i}\left[R^{-1}(\alpha_{n},\beta_{n},\gamma_{n})\vec{\mu}_{I0}(n)\right]\cdot\hat{e}_{j}}{\varepsilon_{0}V\left(E_{0}+\omega-E_{I}\right)},\quad\textrm{ for }\quad i,j=x,y,z.\label{eq:x}
\end{equation}

From now on, we consider the magnetic response of the EMACs to the
external electromagnetic field \cite{k35}.  The magnetic dipole
operator is related to the orbital angular momentum operator $\vec{L}$ of $\pi$
electrons through the relation, $\vec{m} = -e\vec{L}/2m_e$, where $-e$ and $m_e$ 
are respectively the electric charge and mass of the $\pi$ electron. Working
in the molecular coordinate system, one can show that the only
nonvanishing component of the angular momentum operator is along the 
perpendicular direction to the pyridyl plane, i.e. 
\begin{eqnarray}
L_{x} & = &
L_{y}  =0,\\
L_{z} & = & r^{x}p^{y}-r^{y}p^{x}\nonumber \\
 & = & \frac{1}{2}\left(r^{x}p^{y}+p^{y}r^{x}-r^{y}p^{x}-p^{x}r^{y}\right)\nonumber \\
 & = & im_{e}\left(r^{x}H_{0}r^{y}-r^{y}H_{0}r^{x}\right),
\end{eqnarray}
where we used the relation, $p^{\alpha}  =
m_{e}\dot{r}^{\alpha}=im_{e}\left[H_{0},r^{\alpha}\right]$, $\alpha=x,y$ \cite{Bohm89}.
Therefore, the magnetic dipole operator in the molecular coordinate
system is given by 
\begin{eqnarray}
  \vec{m}  & = & \frac{-e}{2m_{e}}L_{z}\hat{e}_{z}\nonumber \\
  & = & \frac{-ie}{2}\hat{e}_{z}\left(r^{x}H_{0}r^{y}-r^{y}H_{0}r^{x}\right)\nonumber \\
  & = & \frac{-ie}{2}\hat{e}_{z}\sum_{k,k^{\prime}} 
  \sum_{k_{1}}E_{k_{1}}\left(r_{kk_{1}}^{x}r_{k_{1}k^{\prime}}^{y}-r_{kk_{1}}^{y}r_{k_{1}k^{\prime}}^{x}\right) |\Psi_k\>\<\Psi_{k^\prime}|,
\end{eqnarray}
where 
\begin{equation}
  r_{II^{\prime}}^{\alpha}=\<\Psi_{I}|r^{\alpha}|\Psi_{I^{\prime}}\>.
\end{equation}
Notice that as $r^{\alpha}$ is a single-electron operator and two
different excited states $|\Psi_{I}\>$ and $|\Psi_{I^{\prime}}\>$
differ from each other by two single-electron states, $|\Psi_I\>$ and $|\Psi_{I^\prime}\>$ 
should not be two different excited states\cite{k34}.

Similarly, since 
\begin{equation}
  \<\Psi_{0}|\vec{m}^{\prime}|\Psi_{0}\> = -\textrm{Re}
  \sum_{I=1}^{18} \frac{\vec{m}_{I0}\cdot\vec{B}_{0}}{E_{0} 
    +\omega-E_{I}}\vec{m}_{0I}e^{-i\omega t},
\end{equation}
the magnetic response is 
\begin{eqnarray}
  \vec{B} & = & \rttensor{\mu}_{0}\vec{H}_{0}+\rttensor{\mu}_{0}\frac{\vec{M}}{V}\nonumber \\
  & = & \rttensor{\mu}_{0}\vec{H}_{0} - \sum_{n=1}^{N}
  \sum_{I=1}^{18}\frac{\mu_{0}\vec{m}_{I0}(n)\cdot\vec{B}_{0}}{V\left(E_{0}+\omega-E_{I}\right)}\vec{m}_{0I}(n) 
  \nonumber\\ 
 & \equiv & \rttensor{\mu}\vec{H}_{0},
\end{eqnarray}
and the relative permeability of the material in the lab coordinate system
is  
\begin{equation}
  \mu_{ij}^{r}\equiv\delta_{ij}-\mu_{0}\sum_{n=1}^{N}\sum_{k=1}^{18} 
  \frac{\left[R^{-1}(\alpha_{n},\beta_{n},\gamma_{n})\vec{m}_{0k}(n)\right]
    \cdot
    \hat{e}_{i}\left[R^{-1}(\alpha_{n},\beta_{n},\gamma_{n})\vec{m}_{k0}(n)\right]\cdot\hat{e}_{j}}
  {V\left(E_{0}+\omega-E_{k}\right)},\label{eq:y} 
\end{equation}
for  $i,j=x,y,z$.

\subsection{Permittivity and Permeability}

As the symmetric center is chosen as the origin of coordinate, the
nuclear contribution can be neglected and thus the electric dipole
moment operator is 
\begin{equation}
\vec{\mu}=-\sum_{j=1}^{6}e\vec{r}_{j},
\end{equation}
where $\vec{r}_{j}$ is the position vector of $j$th $\pi$ electron. 
According to Ref. \cite{k34}, for the case with single-electron operators, 
the matrix elements are given by 
\begin{eqnarray}
\<\Psi_{0}|\vec{\mu}|\Psi_{I}\>=\<\Psi_{0}|\vec{\mu}|\Psi_{a\sigma}^{r\sigma}\>=-e\<\psi_{a}|\vec{r}|\psi_{r}\>=-e\vec{r}_{ar}.
\end{eqnarray}
Here, 
\begin{equation}
\<\psi_{a}|\vec{r}|\psi_{r}\>=\sum_{j=1}^{6}c_{aj}^{*}c_{rj}\vec{R}_{j},
\end{equation}
where $\vec{R}_{j}$ is the position of $j$th site in the lab coordinate
system.

The magnetic dipole moment operator is 
\begin{equation}
\vec{m}=-\frac{ie}{2}\hat{e}_{z}
\sum_{I\neq0}(E_{I}-E_{0})(r_{II}^{x}r_{I0}^{y}-r_{II}^{y}r_{I0}^{x})|\Psi_{I}\>\<\Psi_{0}|+\textrm{h.c.}, 
\end{equation}
where we have used the relation $\vec{r}_{II^{\prime}}=0$ if $I\neq I^{'}$
and both $I$ and $I^{\prime}$ are not equal to zero.

\section{Result and Discussion\label{sec:result}}

\subsection{General Result of Permittivity and Permeability}
\begin{figure}
\includegraphics[clip,scale=0.4]{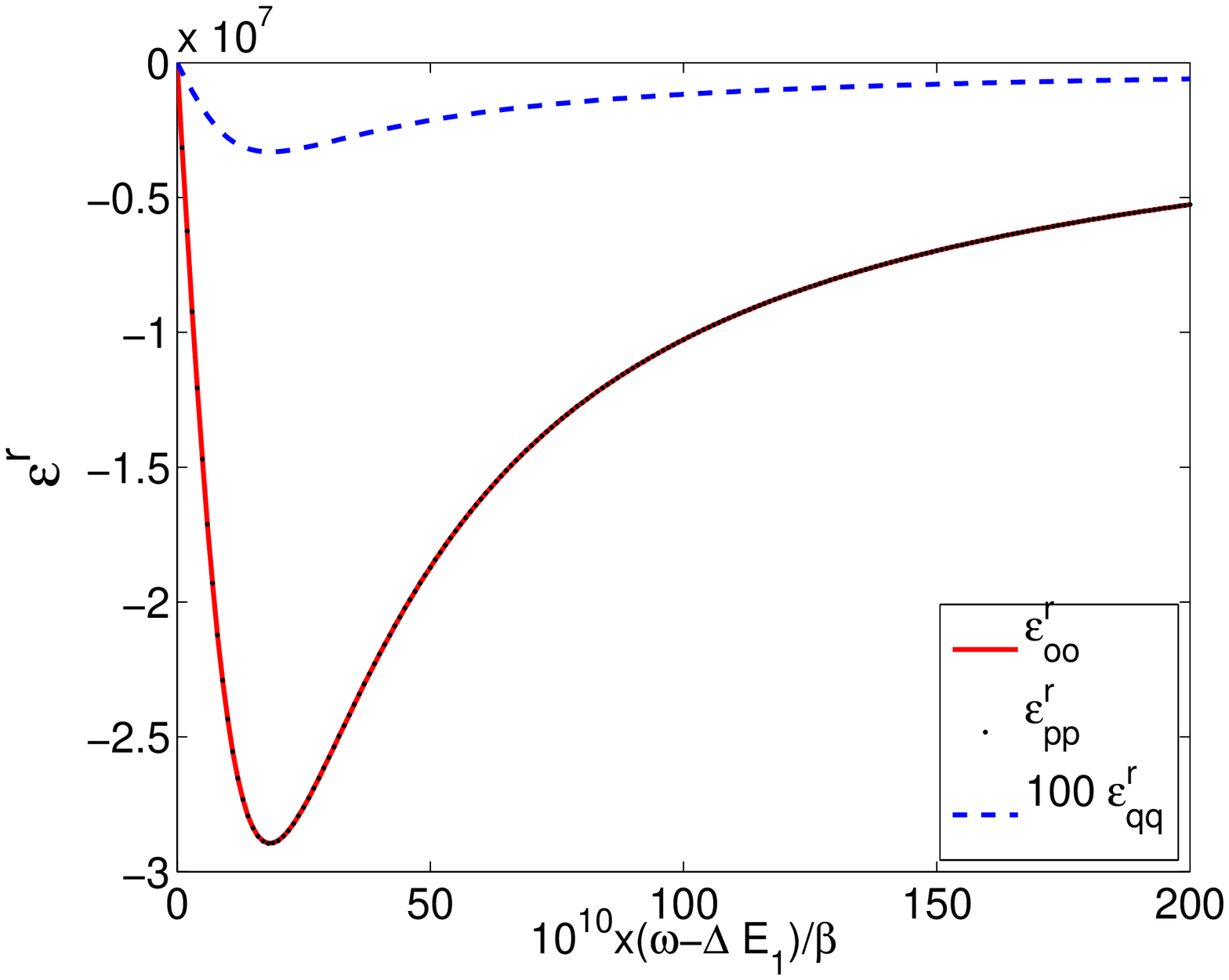}\includegraphics[clip,scale=0.4]{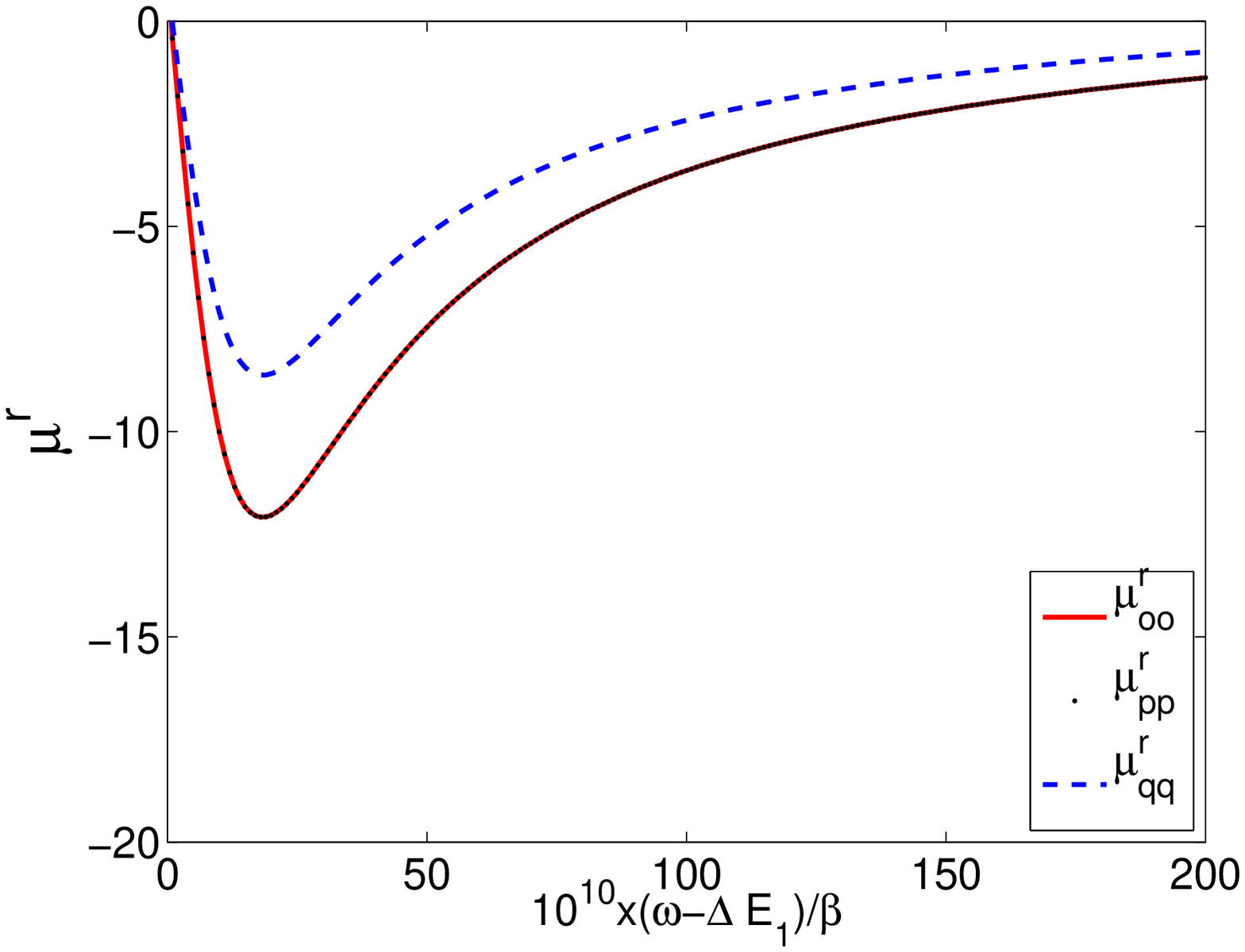}
\caption{The (left) permittivity and (right) permeability of EMACs. The horizontal
and vertical coordinates are two dimensionless quantities. They are
frequency, permittivity and permeability respectively. The three color
lines demonstrate three eigenvalues of permittivity and permeability
tensor. The permittivity and permeability along $o$, $p$ and $q$
directions (see Figure \ref{fig:main}) are all negative here. Thus
we have three dimensional negative permittivity and permeability.
\label{fig:re1}} 
\end{figure}

Equations~(\ref{eq:x}) and (\ref{eq:y}) derived in previous section will be
employed here to show that a material consisting of EMAC complexes can
exhibit negative permittivity and permeability simultaneously in the
UV-Vis region. For anisotropic material in an electromagnetic field
with arbitrary direction, the permittivity and permeability are
generally second rank tensors. In our numerical simulation, we use the
following parameters \cite{k31} for the $\pi$ electrons, i.e., the
site energies $\alpha_{C}=0$, $\alpha_{N}=\alpha_{C}+0.5\beta_{CC}$,
the coupling strengths $\beta_{CC}=-3.6$ eV,
$\beta_{CN}=0.8\beta_{CC}$, and the excited-state life time
$\tau=0.1\,\mu\text{s}$.  And the rotation angles for two successive
pyridyl groups in the smallest trimetal EMACs are $\alpha_{n}=0.105$,
$\beta_{n}=0.829$, $\gamma_{n}=-0.105$.  We rotate the tensors of
$\varepsilon$ and $\mu$ to a special coordinate system, in which
$\varepsilon$ is diagonalized. In this special coordinate system,
$\mu$ has very small off-diagonal terms. Therefore, we only consider
the diagonal terms of $\mu$. The results of computation of electric
permittivity and magnetic permeability tensors along three principal
axes are shown in Figure~\ref{fig:re1}.  Due to the four-fold
rotational symmetry of the EMAC complexes, one principal axis
($q$-axis) of the dielectric permittivity ellipsoid coincides with the
central metal string, while the other two axes ($o$- and $p$-axes)
point to the perpendicular lateral directions.  In addition, because
of the four-fold rotational symmetry of the EMAC complexes, the
dielectric permittivity (magnetic permeability) along the two lateral
axes are equal.  Moreover, it is interesting to note that the
component along the central metal string for the dielectric
permittivity is significantly smaller than the other two components
lying perpendicular to the central metal string, as will be explained in
the next section.  Apparently, the permittivity and permeability of
EMACs are simultaneously negative in all these three principal
directions. As a result, for any given direction, since negative refraction 
will occur in certain frequency window where $\varepsilon$ and $\mu$ 
are simultaneously negative, EMACs are three-dimensional NIM material. 
Generally speaking, in the experimental investigations it is difficult to align 
the molecules in a specific direction. For an arbitrary direction, the permittivity 
and permeability can be approximated by the diagonal terms as long as 
the off-diagonal terms are sufficiently small. In this case, we could observe 
the negative refractive index for both negative diagonal terms of the permittivity 
and permeability.

In Figure~\ref{fig:re1}, we only show the numerical results for the
negative refraction around the first transition frequency, i.e.,
$\omega\sim E_{1}-E_{0}$, where contributions are mainly from the transition
dipole moments, $\mu_{01}$ and $m_{01}$.  In general, the negative
refraction can also occur around other transition frequencies, but the
widths of the negative refraction are much narrower. Besides,
according to our numerical simulation, which is not shown here, the
width of the negative refraction window is very sensitive to the
excited state life time. For a sufficiently small $\tau$, the window
width might be greatly reduced and the negative refraction may even
disappear.

\begin{figure}
\includegraphics[bb=50bp 430bp 550bp 700bp,clip,scale=0.7]{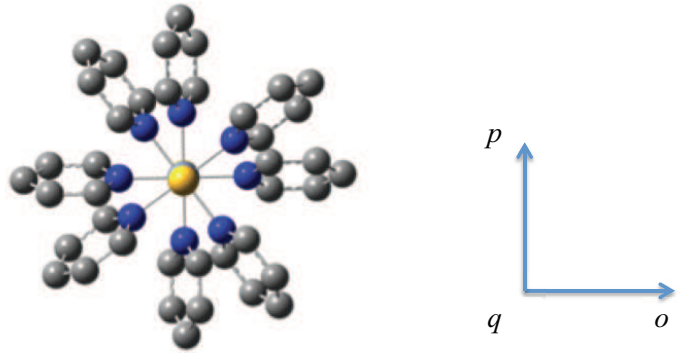}
\caption{The main axes of the permittivity and permeability of EMACs. The $o$,
$p$, $q$ axes in this figure are the main directions of $\varepsilon$
and $\mu$ which are different from those axes in Figure~\ref{fig:two}.
\label{fig:main}}
\end{figure}

\begin{figure}
\includegraphics[bb=0bp 0bp 223bp 153bp,clip,width=12cm]{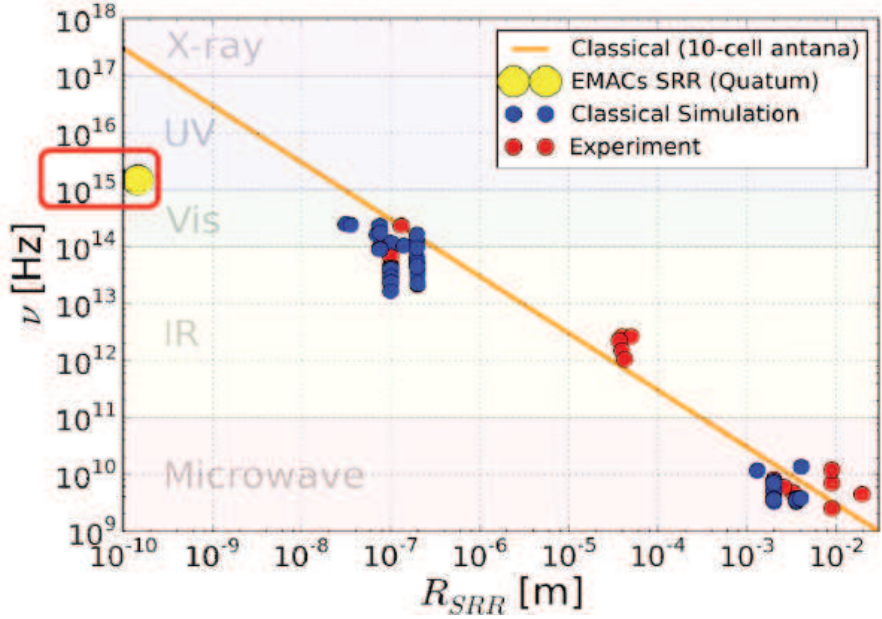} 
\caption{Quantum effect on the response frequency of negative refractive index. The
data for classical materials (blue and red points) in this figure
are supported by references
\cite{k2,p1,p2,p3,p4,p5,p6,p7,p8,p9,p10,p11,p12,p13,p14}
\label{fig:EMACS-3}} 
\end{figure}

We also wish to point out that, unlike the traditional micro-scale
SRRs, quantum confinement in molecular SRRs plays an important role on the
response frequency where negative refraction occurs.  It is known for a
long time that the fundamental wavelength of electromagnetic field
absorbed by a classical metallic particle or attenna with length $L$
is given by $\lambda = 2L$; while the wavelength absorbed by
molecular-scale quasi-metallic objects or unsaturated molecules such
as conjugated polyenes is $\lambda=400 L$, which is much larger than
the length of polyenes due to the quantum confinement
effect.\cite{Platt56,Platt60} It is interesting to note that similar
trend is also observed in the response frequency where negative
refraction occurs by many experimental and theoretical investigations
as shown in Figure~\ref{fig:EMACS-3}. Basically, the classical
metamaterials based on the different sizes of SRRs show a good linear
relationship between response frequency and their dimensions on the
log-log plot.  When the dimension of SRRs is shrunk to the nanometer scale,
by simple extrapolation of the classical linear relationship, we would
expect the negative indexes to occur in the frequency domain far
higher than the UV-Vis frequency, which is of course incorrect.
Instead, the negative indexes of EMAC complexes in the near
ultraviolet regime can only be explained by the quantum confinement
effect. In the case of EMAC complexes, a pyridine molecule is about a
nanometer, whose response frequency is hundreds times smaller than
that predicted by the classical theory. This suggests that for
visible/near-ultraviolet light we could obtain a meta-material of
negative indexes with significantly smaller volume, i.e., $\sim10^{-6}$,
as compared to the classical meta-material.  Therefore, the promising
meta-material made by EMAC complexes would gain much advantage over
its classical counterpart as it can be made sufficiently smaller.

\subsection{PMO Analysis}

In order to understand the structure-property relationship more clearly,
here in this section, we perform the standard perturbation molecular
orbital (PMO) theory.\cite{Michl} We start from the simple H\"uckel
description of a benzene molecule, where all site energies and resonance
integrals are equal, 
\begin{eqnarray}
{\cal H}_{\mathrm{B}} & = &
\alpha\sum_{j=1}^{6} |j\>\<j|
+\beta\sum_{j=1}^{6}\left(
|j\>\<j+1|+|j+1\> \<j|
 \right).
\label{eq:benzene} 
\end{eqnarray}
Here $N=6$ is the total number of sites in a benzene molecule.
Extension to other $N(=4n+2)$-annulenes can be done straightforwardly.
Taking a discrete Fourier transform
$|k\>=\frac{1}{\sqrt{6}} \sum_{j=1}^{6}e^{-ikj}|j\>$, 
$k={m \pi
  }/{3}$ and $m=-3,-2,\cdots,2$, ${\cal H}_{\mathrm{B}}$ can be diagonalized
as ${\cal H}_{\text{B}} = \sum_k \varepsilon_k^{(0)} |k\>\<k|$, where
$\varepsilon_k^{(0)} = \alpha + 2 \beta \cos k$. 
In the case of a benzene molecule, $k_{1}=-\pi$ and $k_{4}=0$
correspond to the lowest and the highest molecular orbitals,
respectively, while $k_{2,6}=\mp2\pi/3$ and $k_{3,5}=\mp\pi/3$
correspond to two sets of degenerate frontier molecular orbitals, respectively.

In the PMO analysis, a pyridine molecule is analyzed by making a chemical
perturbation on its parent molecule, i.e. a benzene. Within the H\"uckel
framework, replacing a carbon atom by a nitrogen atom at site $1$
is equivalent to introducing a perturbation term, i.e. ${\cal H}_{\text{P}} =
{\cal H}_{\text{B}} + {\cal V}$, where 
\begin{equation}
\begin{split}
{\cal V} & =\Delta\alpha |1\>\<1|
+\Delta\beta\left(
|1\>\<2|+|6\>\<1|
\right)+\mathrm{h.c.}
\end{split}
\label{eq:2}
\end{equation}
Transforming to the delocalized orbital representation through the
discrete Fourier transform, we get 
\begin{equation}
{\cal H}_{\mathrm{P}}=\sum_{k} (\varepsilon_{k}^{(0)} + \delta \varepsilon_k  )|k\>\<k|
+ \sum_{k<k^{\prime}}J_{kk^{\prime}} |k\>\<k'| +\mathrm{h.c.},
\end{equation}
where the diagonal and off-diagonal
perturbations terms are related to original chemical perturbations
$\Delta \alpha$ and $\Delta \beta$ by 
\begin{eqnarray}
\delta\varepsilon_{k} & = &
\frac{\Delta \alpha}{6} +\frac{2\Delta \beta}{3}\cos k,\\
J_{kk^{\prime}} & = & \frac{\Delta\alpha}{N}\cos(k-k^{\prime})+\frac{2\Delta\beta}{N}\left[\cos(k-2k^{\prime})+\cos(kN-k^{\prime})\right].
\end{eqnarray}
and they correspond to the first-order and higher-order perturbation corrections to the energies of
molecular orbital energies, respectively. Note that the degeneracies
are not removed by the diagonal terms, i.e. $\delta
\varepsilon_2 = \delta \varepsilon_6 $ and $\delta \varepsilon_3 =
\delta\varepsilon_5$. 
Incorporating the couplings between the degenerate molecular orbitals,
the unperturbed Hamiltonian can be analytically diagonalized as 
\begin{eqnarray}
  {\cal H}_{\text{P}} & = & \sum_{k}\varepsilon_{k}|k\>\<k|
  +J_{k_{2}k_{6}}|k_2\>\<k_6| + J_{k_{3}k_{5}} |k_3\>\<k_5| +\mathrm{h.c.}\nonumber \\
  & = & \sum_{k}\varepsilon_{k}^{\prime} |\tilde k\>\<\tilde k|,
\end{eqnarray}
where $\varepsilon_{k}=\varepsilon_{k}^{(0)}+\delta \varepsilon_{k}$, for $k=k_{1},k_{4}$, $\varepsilon_{k}^{\prime}=\varepsilon_{k}$
and $|\tilde k\> = |k\>$, for other $k$'s, we have
$\varepsilon_{k_{2}}^{\prime},  
\varepsilon_{k_{6}}^{\prime}  =  \varepsilon_{k_{2}} \mp
J_{k_{2}k_{6}}$,  $\varepsilon_{k_{3}}^{\prime}, 
\varepsilon_{k_{5}}^{\prime}  =  \varepsilon_{k_{3}}\mp
J_{k_{3}k_{5}}$, 
and $|\tilde k_2\>, |\tilde k_6\> = (\mp |k_2\> +|k_6\>)/\sqrt{2}$, $|\tilde k_3\>,
|\tilde k_5\> = (\mp |k_3\> + |k_5\>)/\sqrt{2}$.
The couplings between two degenerate pairs are respectively $J_{k_2k_6} = (2\Delta
\beta - \Delta \alpha)/6 = -0.1 \beta_{\text{CC}}/12$ 
and $J_{k_3k_5}
= -(2\Delta \beta+\Delta \alpha)/12 = -0.9 \beta_{\text{CC}}/12$. 
The matrix elements of dipole operator for the transition to the
lowest excited state can also be evaluated
straightforwardly 
\begin{eqnarray}
\<\Psi_1|\vec{\mu}|\Psi_0\>  \propto  \left<\tilde k_3\right|\vec{r}\left|\tilde k_6\right>
 & = & \frac{\sqrt{3}i}{2}(\vec{R}_{1}+\vec{R}_{2}),
\end{eqnarray}
where we have made use of $\vec{R}_{j+3}=-\vec{R}_{j}$. Because  $\left<\tilde k_3\right|\vec{r}\left|\tilde k_6\right>$ has a greater component in the $x$ direction than in the $y$ direction, we would expect $\varepsilon_{oo(pp)}^r$ to be sufficiently larger than $\varepsilon_{qq}^r$.

On the other hand, due to the symmetry of the unperturbed Hamiltonian
$\mathcal{H}_{B}$, the diagonal matrix elements of the displacement
operator vanish, i.e. 
\begin{equation}
\left<k\right|\vec{r}\left|k\right>=0.
\end{equation}
As a result, all $\<\psi_{k}|\vec{r}|\psi_{k}\>$'s disappear to the
present order, and their contribution originates from the next order
due to the presence of ${\cal V}$. And it also suggests that
due to the presence of nitrogen atom, the one-atom-substituted H\"uckel model
makes an SSR and thus can interact with the applied magnetic
field. 

Besides, we also remark that the molecular SSR can be realized by 
molecules other than EMACs. For example, (P)-2,2'-biphenol~\cite{Mineo12} is two connected 
phenol where a hydrogen atom of benzene is replaced by a hydroxyl group.  
In this molecule, due to the presence of a hydroxyl group, the site energy and 
two coupling constants of the connected carbon atom are modified. Furthermore, 
since the two phenol are not in the same plane, this molecule could response 
to the eletromagnetic field in all three dimensions. Another possible candidate is  
2,5-dichloropyrazine~\cite{Kanno10} where two nitrogen atoms are substituted for two carbon 
atoms and two hydrogen atoms are replaced by two chlorine atoms. 
In this molecule, due to this chemical tuning, the energy spectrum of the excited 
states is significantly adjusted. Therefore, our proposal may work not only in EMACs 
but also in molecules with a SRR configuration.

\section{CONCLUSION\label{sec:CONCLUSION}}

In this paper, we discover a new family of conjugated molecules that
have the potential to exhibit negative refractive index in the UV-Vis
region. The molecules of interest are single-nitrogen-substituted
heterocyclic annulenes, in which heteroatoms serves as splits to turn
on magnetic resonance as their classical analogue. With heterocyclic
annulenes as molecular split-ring resonators (SSRs), we show that extended
metal atom chains (EMACs) can realize negative permittivity and permeability
simultaneously in quantum regime. A simple tight-binding calculation
based on linear response within the electric and magnetic dipole approximations
gives a reasonable estimate of the negative index region of the
three dimensional and helical arrangement of these molecular SSRs. 
Although we have adopted a simplified independent electron
model, the Coulombic interaction between electrons should not destroy
the prediction of negative refraction. Furthermore, besides EMACs,
a family of single-nitrogen-substituted heterocyclic annulenes may
also serve as the negative refraction materials if the molecules are
appropriately arranged.


\section{Acknowledgement}

The research was supported by the National Science Council, Taiwan
and the Center of Theoretical Sciences of National Taiwan University.

\newpage{}
\textbf{TOC Graphic.}
\includegraphics[bb=0bp 0bp 175bp 138bp,width=5cm,height=5cm]{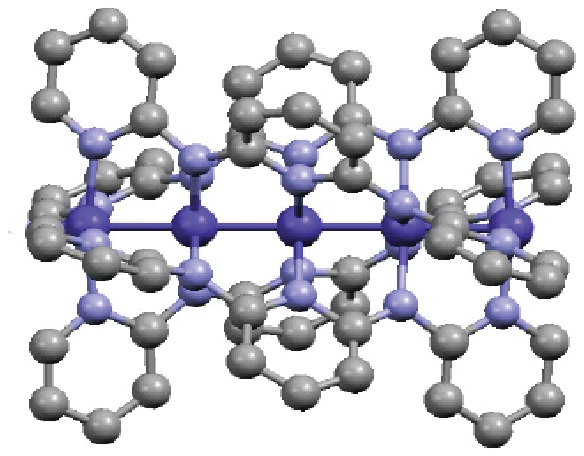}

\end{document}